\documentclass[12pt]{article}

\usepackage{amsmath}
\usepackage{amsfonts}
\usepackage{latexsym}
\usepackage[mathscr]{euscript}

\renewcommand{\theequation}{\thesection.\arabic{equation}}

 \def\scr{\mathscr}

  \def\D{{\scr D}}
  \def\Db{\bar\D}

  \def\L{{\scr L}}
  \def\Lb{\bar\L}
  \def\C{{\cal C}}

\newcommand{\N}{{\mathscr N}}

\def\e{{\rm e}}

\def\l{\langle}
\def\r{\rangle}
\def\pr{\partial}

\newcommand{\half}{{\textstyle \frac{1}{2}}}
\def\quar{{\textstyle \frac{1}{4}}}

\newcommand{\Leff}{\L_{\rm eff}}
\newcommand{\Leffb}{\Lb_{\rm eff}}

\newcommand{\Geff}{\Gamma_{\rm eff}}

\newcommand{\dS}{\!\!{\rm d}^6z\,}
\newcommand{\dSb}{\!\!{\rm d}^6\bar z\,}
\newcommand{\dV}{\!\!{\rm d}^8z\,}

\newcommand{\Ixi}{I_{\xi}}

\newcommand{\IPk}{I_{\Phi}}
\newcommand{\Igf}{I_{\rm gf}}

\newcommand{\LPk}{\L_{\Phi}}
\newcommand{\Lgf}{\L_{\rm gf}}
\newcommand{\Lgfb}{\L_{\rm gf}'}
\newcommand{\hM}{\widehat{M^2}}

\newcommand{\Lg}{\L_{g}}

\newcommand{\Lxi}{\L_{\xi}}
\newcommand{\Lxib}{\L_{\xi}'}

\newcommand{\ug}{u_{g}}

\newcommand{\uxib}{u_{\xi}'}

\newcommand{\Pc}{(\Db^2+R)}
\newcommand{\Pac}{(\D^2+\bar R)}

\newcommand{\W}[2]{ {}^{#1}_{#2}\!W{} }

\def\phys{|{\rm p}\rangle}
\def\dphys{|\delta {\rm p}\rangle}
\def\lphys{\langle{\rm p}'|}
\def\ldphys{\langle\delta {\rm p}'|}

\newcommand{\al}{\alpha}
\def\l{\langle}
\def\r{\rangle}
\def\da{{\dot\alpha}}
\def\be{\beta}
\def\db{{\dot\beta}}

\def\de{\delta}

\def\is{{^{\!(\sigma)}}}
\def\il{{^{\!(\Lambda)}}}
\def\de{\delta}
\def\si{\sigma}
\def\ga{\gamma}

\newcommand{\eps}{\varepsilon}

\newcommand{\tfr}[2]{{\textstyle \frac{#1}{#2}}}

\newcommand{\fdq}[2]{\frac{\delta #1}{\delta #2}}

\def\ts{\textstyle}

\newcommand{\dx}{\!\!{\rm d}^4x\,\,}

\def\tP{{\tilde \Phi}}
\def\hP{{\hat \Phi}}

\begin{document}
\vspace*{-5mm}
\begin{flushright}
hep-th/9907169\\ NTZ 14/1999\\ July 1999
\end{flushright}

\thispagestyle{empty}

\vspace{-0.5cm}
\begin{center}
{\Large 
 \bf Superconformal Transformation Properties of }

{\Large \bf the Supercurrent II: Abelian Gauge Theories}

\vspace{1cm}

{\parindent0cm
Johanna Erdmenger\footnote{Supported by Deutsche Forschungsgemeinschaft,
e-mail: Johanna.Erdmenger@itp.uni-leipzig.de} \\
Christian Rupp\footnote{Supported by Graduiertenkolleg
 ``Quantenfeldtheorie: Mathematische Struktur und physikalische
Anwendungen",
e-mail: Christian.Rupp@itp.uni-leipzig.de} \\
Klaus Sibold\footnote{e-mail: sibold@physik.uni-leipzig.de}
}

\vspace{0.7cm}

Institut f{\"u}r Theoretische Physik\\
Universit{\"a}t Leipzig\\
Augustusplatz 10/11\\
D - 04109 Leipzig\\
Germany
\end{center}

\vspace{1.5cm}

\centerline{\small \bf Abstract}\vspace*{-2mm} { \small \noindent 

We derive the superconformal transformation properties of the
supercurrent for $N=1$ supersymmetric QED in four dimensions within the
superfield formalism. 
Superconformal Ward identities for Green functions involving
insertions of the supercurrent are conveniently derived 
by coupling the supercurrent to the appropriate prepotential
of a classical curved superspace background, and
by combining superdiffeomorphisms and super Weyl transformations.
We determine all superconformal anomalies of SQED on curved superspace
within an all-order perturbative approach and derive a local
Callan-Symanzik equation. Particular importance is
given to the issue of gauge invariance. }

\begin{tabbing}
PACS numbers: \= 04.62+v, 11.10Gh, 11.30Pb, 11.15.-q.\\ Keywords:\>
Quantum Field Theory, Superconformal Symmetry, Gauge Theory,
\\ \> Curved Superspace
Background, Supercurrent, Anomalies.
\end{tabbing}
\newpage

\section{Introduction}

Conformal symmetry and superconformal symmetry in particular
have attracted much attention within four-dimensional quantum field theory for
a long period of time as well as very recently. One of the numerous
interesting aspects of conformal symmetry is its interplay with
renormalisation. For quantised field theories, conformal symmetry is
broken by anomalies in general. The exact determination of these
conformal anomalies provides detailed information about the
renormalisation behaviour of a given quantum field theory. 
In particular it is of interest to study the superconformal
transformation properties of the supercurrent, an axial vector
superfield from which all currents of the superconformal group may be
obtained by means of a moment construction. The exact knowledge
of this transformation behaviour is expected to be of relevance for instance
for a potential proof of the Zamolochikov C theorem in four dimensions.
In this context it is essential to study gauge theories, since 
for several examples of $N=1$ supersymmetric non-abelian gauge theories 
it has been shown that the coefficient of the topological
Euler density is larger in the UV than in the IR limit, in agreement with
the conjectured extension of the C theorem \cite{Anselmi}.

Here we consider field theories away from fixed points, where
anomalies involving the dynamical fields may be present.  
In a recent series of papers \cite{ERS1,ERS2,ERS3} 
we have determined the superconformal transformation behaviour of the
supercurrent for the massless Wess-Zumino model. For deriving superconformal
Ward identities for correlation functions involving the supercurrent
we have coupled the supercurrent to the appropriate prepotential of
a classical curved superspace background. On curved superspace,
superconformal transformations are given by the combination of
superdiffeomorphisms and super Weyl transformations. For theories with
a conserved energy-momentum tensor, superdiffeomorphisms are
anomaly-free, such that
the breakdown of conformal symmetry upon quantisation is entirely
determined by the anomalies of Weyl symmetry. These manifest
themselves in anomalous contributions to the Weyl symmetry Ward
identity. In \cite{ERS1} we have determined all dynamical anomalies in
this  Ward identity. Since all dynamical anomalies are
given by terms which involve the usual $\beta,\gamma$ functions as
their coefficients, it  may be
interpreted as a {\it local} Callan-Symanzik (CS) equation. The integral of
this equation agrees with the well-known global CS equation.
Furthermore there are purely geometrical superconformal anomalies
which involve only the external classical super\-gravity fields
\cite{ERS2}. These lead to purely local contributions to Ward identities
for Green functions with multiple insertions of the supercurrent. 
Such Ward identities have been discussed in \cite{ERS3}, paying careful
attention to the  
renormalisation behaviour of Green functions with multiple insertions
of composite operators. All these results are valid in a perturbative
approach to all orders in $\hbar$.

In the present paper we extend this analysis to the more involved
case of $N=1$ supersymmetric QED with massless matter and massive
vector fields. Superconformal Ward identities are
again derived by considering  \pagebreak a classical curved space
background\footnote{To lowest order in the supergravity prepotentials,
the geometrical method for deriving such Ward identities has been
discussed in \cite{West}.}. 
However in this case the compatibility of superconformal
transformations and of gauge transformations has to be ensured.
For renormalisation we  use the BPHZ approach, and we follow the
procedure
developed in \cite{PS1,PS2} for abelian gauge theories on flat space,
i.e.~for the superconformal transformations properties of Green 
functions without insertions of the supercurrent.
 
For defining a covariant gauge field propagator, we fix the gauge explicitly, 
such that gauge invariance is broken. 
The local Callan-Symanzik equation and the conformal transformation
properties of the supercurrent may then be derived in close analogy to
the Wess-Zumino model. 
Since the gauge fixing terms are not conformally invariant, they
contribute to the final results. However we are able to show that
these terms are {un\-physi\-cal} in the sense that they do not contribute
when acting on states in the physical Hilbert space. As far as the
superconformal transformation of the supercurrent is concerned, we
show that some of the gauge non-invariant terms cancel with the
superconformal transformations of the physical states themselves.

This supersymmetric 
renormalisation scheme appears to be well-suited for deriving the
transformation properties of composite operators to all orders in
$\hbar$, and extends in a natural way to the classical curved space
background. A different approach to studying supersymmetric 
composite operators which
keeps gauge invariance manifest has been discussed in \cite{Suzuki} on
flat space. This has also been applied to anomalies at the one-loop level.

This paper is organised as follows.  We begin by discussing the
superconformal and gauge transformation properties of the dynamical
and the background fields in section 2. Using these results we discuss
the superconformal and gauge transformation properties of SQED on
curved superspace in section 3.  In particular we derive a local CS
equation which characterises all dynamical superconformal anomalies in
terms of the usual $\beta$ and $\gamma$ functions. With the help of
this equation we derive the superconformal transformation properties
of the supercurrent on flat space. In section 4 we address the issue
of gauge invariance. Section 5 contains some concluding remarks.

\section{Symmetry transformations of the fields}
\subsection{Superconformal Transformations}

\setcounter{equation}{0}

We consider curved superspace in an approach to supergravity
characterised by an axial vector superfield $H$,
\begin{gather}
H = H^a \pr_a \,
\end{gather}
and a chiral compensator $\phi \equiv \exp(J)$ satisfying
\begin{gather}
\bar D_\da \phi = 0 \, , \qquad D_\al \bar \phi = 0 \,,
\end{gather}
where
\begin{gather} \label{basis}
D_A \equiv (\pr_a, D_\al, \bar D^\da )
\end{gather}
are the well known partial and flat space supersymmetry derivatives
respectively, which span the tangent space\footnote{In our conventions 
the metric has diagonal elements $(+1,-1,-1,-1)$.The covariant derivatives
satisfy $\{ D_\al, \bar D_\da\} = 2 i \sigma^a_{\al \da} \pr_a$.}.
This approach to supergravity is presented in \cite{ERS1},
\cite{ERS2}, \cite{ERS3}. Further details may be found in the textbooks
\cite{bk}, \cite{ggrs}. This applies also to the definition the supercovariant
derivatives, denoted by
\begin{gather}
\D_A \equiv ( \D_a, \D_\al, \bar \D^{\da} ) \, ,
\end{gather}
which are the covariant derivatives associated
to the group of superdiffeomorphisms on curved superspace and which
depend on $H$. Moreover it should be noted that 
throughout this paper we use the curved space
chiral representation
which is analogous to the flat space chiral representation in which all fields
$\tilde \Phi$ are replaced by 
$\Phi = \e^{i\theta\sigma^a\bar\theta\partial_a} \tilde\Phi$.
Complex conjugation of $\Phi$ in the chiral representation yields $\bar \Phi$
in the antichiral representation. On curved space, the chiral 
representation expression for the conjugate of $\Phi$ 
is given by  ${\rm e}^{2iH} \bar \Phi$. This is of particular
importance for the gauge superfield.

The superdiffeomorphisms are given by a complex superfield
$\Lambda$ and its conjugate $\bar \Lambda$, 
\begin{align}
\Lambda &= \Lambda^a \pr_a + \Lambda^\alpha D_\alpha + \Lambda_\da \bar D^\da 
+ \Lambda^{\al \beta} M_{\alpha \beta} + \Lambda^{\da \db} M_{\da \db} \, ,
\nonumber\\
\bar \Lambda &= \bar \Lambda^a \pr_a + \bar \Lambda^\alpha D_\alpha + \bar 
\Lambda_\da \bar D^\da 
+ \bar \Lambda^{\al \beta} M_{\alpha \beta} + 
\bar \Lambda^{\da \db} M_{\da \db} \, , \label{diffeom}
\end{align}
where $\Lambda$ is in the chiral and $\bar \Lambda$ in the 
antichiral representation, and $M, \bar M$ are the gene\-rators of
infinitesimal Lorentz transformations. Covariance constraints imply
that the diffeomorphism components may be written as
\begin{equation}
\Lambda^{\alpha\dot\alpha} = i \bar D^{\dot\alpha} \Omega^\alpha\, ,\;\;\;\;
\Lambda^\alpha = \tfr{1}{4} \bar D^2 \Omega^\alpha \, , \quad
\Lambda_{\dot\alpha}= \e^{2iH} \bar \Lambda_{\dot\alpha} \, , 
\;\;\;\;
\Lambda_{\alpha\beta} = \e^{2iH} \bar \Lambda_{\alpha\beta}\, ,
\label{omegadef}
\end{equation}
in terms of an unconstrained superfield $\Omega$, with corresponding
expressions for $\bar \Lambda$.
The diffeomorphism transformation properties of the supergravity
prepotentials 
are given by
\begin{gather} \label{Htrans}
\e^{2iH}  \longrightarrow  \e^\Lambda \e^{2iH} \e^{-\bar \Lambda} \, ,
\end{gather}
for the real axial vector prepotential $H$ and by
\begin{gather} \label{phi2}
\phi^3 \rightarrow  \phi^3 \, \e^{\overleftarrow{\Lambda_c}} \, , \qquad
\Lambda_c = \Lambda^a \pr_a + \Lambda^\al D_\al 
\end{gather}
for the chiral compensator, both in the chiral representation.

The super Weyl transformations are given by
superfields $\sigma$, $\bar \sigma$ which satisfy $\bar D_\da \sigma = 0$,
$ D_\al \bar \sigma = 0$. For the Weyl transformation properties
of the prepotentials we have
\begin{gather} \label{W2}
H \longrightarrow H , \qquad
\phi \longrightarrow \e^\sigma \phi , \qquad 
\bar \phi \longrightarrow \e^{\bar \sigma} \bar \phi.
\end{gather}
It is crucial to note that $H$ is Weyl invariant.
Superconformal transformations satisfy the additional relations
\begin{gather} \label{sigmaomega}
\Lambda = \bar \Lambda \, \Rightarrow \bar D^\da \Omega^\al = D^\al
\bar \Omega^\da \, , \quad \sigma = - {\ts \frac{1}{12}} \bar D^2 D^\al
\Omega_\al \, . 
\end{gather}

For the dynamical fields of supersymmetric QED we have two chiral
matter fields $A_+, A_-$ and their chiral partners, 
which under parity transform as
\begin{gather}
{{\cal P}} \, A_+ = \bar A_-\,, \qquad {{\cal P}} A_- = \bar A_+ \, .
\end{gather}
Their diffeomorphism and Weyl transformation properties are given by
\begin{gather}
A_\pm \longrightarrow \e^\Lambda A_\pm \, , \quad A_\pm \longrightarrow
\e^{-\si} A_\pm \, .
\end{gather}
The spinorial components $\psi_+^\al$, $\psi_-^\al$  of $A_+$, $A_-$ 
form a Dirac spinor
\begin{equation}
\Psi_D = \left( \hspace{-0.5em}\begin{array}{c}\psi_{+\al}\\[1ex]\bar \psi_-^\da
  \end{array}\hspace{-0.5em}
\right)\,,
\end{equation}
such that a theory involving $A_+$, $A_-$ 
describes the coupling of a Dirac particle to the electromagnetic field.
The appropriate gauge superfield $\Phi$ is dimensionless and
real. Under
diffeomorphisms
\begin{gather}
\Phi \longrightarrow \e^\Lambda \Phi \, .
\end{gather}
Since its canonical dimension is zero, $\Phi$ is Weyl invariant in the
classical case discussed here.

It should be noted that in the chiral representation used above, the
reality condition on $\Phi$ reads
\begin{gather}
\Phi = {\rm e}^{2iH} \bar \Phi \, .
\end{gather}
In order to avoid such $H$ dependent constraint on $\Phi$, it is more
convenient to use the real representation in which
\begin{equation} 
\tilde \Phi = {\rm e}^{- i H} \Phi\,,
\end{equation}
such that $\tilde \Phi = \bar{\tilde \Phi}$. By explicit calculation
we find that in the real representation, to lowest order in $H$, 
 the gauge field has the
diffeomorphism transformation property
\begin{gather}
\tP \longrightarrow {\rm e}^{\left(1/2(\Lambda^a + \bar \Lambda^a)
  \pr_a +
  \Lambda^\al D_\al + \bar \Lambda_\da \bar D^\da \right)} \tP.
\end{gather}

\subsection{Gauge transformations}

The gauge transformations of the matter fields are given by
\begin{equation}
\delta_\lambda A_+(z) = -i g \lambda(z) A_+(z)\,, \qquad \delta_\lambda
 A_-(z) = +i g
\lambda(z) A_-(z)\,, \quad \bar\D^\da \lambda = 0 \, ,
\end{equation}
with $g$ the charge and $\lambda$ chiral. For the gauge field we have
\begin{equation}
\delta_\lambda \tP = i \left( \e^{-i H}\lambda - \e^{i H} 
\bar \lambda \right) \, .
\end{equation}
Again it is not convenient to consider such $H$ dependent gauge
transformation. Therefore we use the `gauge flat' representation of
\cite{ggrs},
\begin{equation} 
\tP = \left( \cosh (-i H) + \frac{\sinh(-i H)}{-i H} \circ \half
  H^{\al\da} [D_\al, \bar D_\da] \right) \hP \, \label{gaugeflat}
\end{equation}
for which
\begin{equation}
\delta_\lambda \hP = i \left( \lambda - \bar \lambda \right)\, .
\end{equation}
$\hat \Phi$ is real, 
$\hat \Phi = \bar {\hat\Phi}$.

Moreover the supergravity fields are gauge invariant. 
The infinitesimal forms for all field transformations discussed may be found
in appendix \ref{infinitesimal}. These are relevant for deriving Ward
identities. 

\section{Transformation properties of the supercurrent}
\setcounter{equation}{0}

Our aim is to derive superconformal Ward identities for Green
functions involving insertions of the supercurrent for supersymmetric
quantum electrodynamics (SQED). For renormalisation we use the BPHZ
approach \cite{bphz}, in which integrands whose integrals are
potentially divergent are expanded into power series in external
momenta. Then terms of this series are subtracted such that the
integrals over the remaining terms are well-defined finite expressions. 

We determine the dynamical 
anomalies present in the Weyl symmetry Ward identity to all orders in
perturbation theory. The purely geometrical anomalies depending on the
classical background supergravity fields only are the same as
discussed in \cite{ERS2}, save for the coefficients which are model
dependent.

The transformation properties of the supercurrent $V_a$
are obtained by virtue of the action principle, according to which the
variation of the vertex functional with respect to the external field
$H_a$ yields an insertion of the supercurrent,
\begin{gather} \label{H}
\left[ V_{a} \right] \cdot \Gamma = \, 8 \, \frac{\de}{\de H^a} \Gamma
\, .
\end{gather}
As shown in \cite{ERS1,ERS3}, superconformal Ward identities are
conveniently obtained by com\-bining superdiffeomorphism and super Weyl
transformations on curved superspace. The diffeomorphism invariance of
SQED is preserved by the renormalisation scheme
to all orders in perturbation theory, which ensures energy-momentum
conservation. The anomalies of superconformal symmetry are thus given
by the anomalies of super Weyl symmetry.

The use of the field $\hP$ as defined implicitly in (\ref{gaugeflat})
is vital for ensuring that - discarding gauge fixing terms - 
$V_a$ as given by (\ref{H}) is gauge
invariant. Furthermore the use of $\hP$ ensures that at the classical
level where no anomalies are present, 
$V_a$ is a conformally covariant, i.e.~quasi-primary field. 
This is due to the fact that the Ward operator
\begin{gather}
W(\hP) \equiv \int\dV \Omega^\al{}_{\!\! \rm conf} \, w^\il_\al (\hP) + c.c. 
\end{gather}
with $w^\il_\al(\hP)$ as in appendix \ref{localWardoperators}
and with $\Omega^\al{}_{\!\! \rm conf}$ satisfying (\ref{sigmaomega})
has no $H$ dependent contribution to first order in $H$.

\subsection{SQED on curved superspace}

We discuss the quantised theory within perturbation theory to all
orders in $\hbar$.
A basis of diffeomorphism invariant field monomials involving the chiral 
matter superfields
$A_+$, $A_-$, as well as the real abelian gauge superfield field $\Phi$,
is given by 
\begin{align} 
\IPk &= \int\dS \phi^3 F^\al F_\al \, ,
& \Igf &= \int \dV E^{-1} \Pc \Phi \Pac \Phi \, ,\nonumber \\
I_g &= \int \dV E^{-1} \left( \bar A_+ \e^{g\Phi}A_+ + \bar A_-
  \e^{-g\Phi}A_- \right) \, ,
& \Ixi &= \int \dS \phi^3 R A_+ A_- \, ,\nonumber \\
I_M &= \int \dV E^{-1} \Phi^2 \, ,
& I_m &= \int\dS \phi^3 A_+A_- \, , \label{I} 
\end{align}
where $F_\al = \Pc \D_\al \Phi$ is the field strength, with $R$ the
chiral supersymmetric curvature scalar. $E^{-1}$ is the
general curved superspace integration measure and $\phi^3$, the cube
of the chiral compensator, is the integration measure for chiral subspace.
$\Pc$ and $\Pac$ are chiral and anti-chiral 
projection operators respectively. Furthermore there is also a basis
for local chiral field monomials leading to (\ref{I}) upon
integration, which is given by
\begin{align}
\LPk &= \phi^3 F^\al F_\al \, ,\nonumber \\
\Lgf&= \phi^3 \Pc \left( \Phi \Pc\Pac \Phi\right)  ,
& \Lgfb&= \phi^3\Pc\left(\Phi \Pac\Pc\Phi\right)  ,\nonumber \\
\L_g &= \phi^3 \Pc \left( \bar A_+ \e^{g\Phi}A_+ + \bar
  A_-\e^{-g\Phi}A_-\right) \, ,\nonumber \\
\Lxi &= \phi^3 R A_+ A_- \,,
& \Lxib &= \phi^3 \Pc \left( \bar A_+ \bar A_-\right) \, ,\nonumber\\
\L_M&= \phi^3 \Pc \Phi^2  \, ,
& \L_m&= \phi^3 A_+ A_- \, . \label{L}
\end{align}

With the basis (\ref{I}) the effective action (in the sense of
Zimmermann) for SQED is given by
\begin{equation} \label{Geff}
\Geff = -\tfr{1}{128} z_\Phi \IPk + \tfr{1}{16} z_g I_g
+ \tfr{1}{8} \hat\xi (\Ixi+\bar\Ixi)
+ \quar m(s-1) (I_m+ \bar I_m) + \tfr{1}{16} \hM I_M -\tfr{1}{128\al}\Igf 
\, .  
\end{equation}
Here the coefficients $z_\Phi$, $z_g$, $\hat \xi$ and $\hM$ are power
series in $\hbar$. 
$\IPk$ and $I_g$ are the curved superspace generalisations of the
usual terms contributing to SQED on flat space. $\Ixi$ describes the
coupling of the curved space background to the matter fields. 
The factor $(s-1)$ in front of the matter field mass terms is an
auxiliary parameter participating in the subtractions, which is
necessary for consistent renormalisation of massless theories, as
described below. It may be set to $s=1$ at the very end of the
renormalisation procedure. We also include a mass term $\hM I_M$ for
the gauge field. This ensures in particular that
there is no  infrared problem. Finally for a
consistent perturbative approach which requires a well-defined
covariant gauge field propagator, we fix the gauge by including the
gauge fixing term  $\Igf$ with $\al$ the gauge parameter.
Therefore the gauge Ward identity has the form
\begin{align}
w^{(\lambda)}(A,\hP) \Gamma & = i\, \phi^3 \Pc \left\{ -\tfr{1}{128\al} R \Pac\Phi
  -\tfr{1}{128\al} \Pac\Pc\Phi + \quar \hM \Phi \right\} \, ,
\end{align}
where the local gauge Ward operator is given by
\begin{align}
w^{(\lambda)}(A,\hP) &=  ig \left( A_- \frac{\de}{\de A_-} - A_+
\frac{\de}{\de A_+} \right) \, + \, i \, \bar D^2 \frac{\de}{\de \hP}
\, .
\end{align}

\subsection{Dynamical anomalies}

We derive a local Callan-Symanzik equation in close analogy to the
procedure followed for the Wess-Zumino model in \cite{ERS1}. 
According to the discussion of the Weyl transformation proper\-ties of
both
dynamical and geometrical fields in section 2,
the local chiral Weyl symmetry Ward operator is given by
\begin{align} \label{ww}
w^{(\si)} (A,J)  &=  \frac{\de}{\de J}  - A_+ \frac{\de}{\de A_+} -
A_-\frac{\de}{\de A_-} \, ,
\end{align}
where $J$ is given by $\phi \equiv \exp(J)$ with $\phi$ the chiral
compensator.
Applying this operator to the vertex functional $\Gamma$ as given by
(\ref{Geff}) we obtain
\begin{align} \label{wweyl}
w^{(\si)} (A,J) \Gamma & = -\tfr{3}{2}\, [S]_3^3\cdot \Gamma \,,\\
S &=  {\ts \frac{1}{16}} \hM \L_M   
+ {\ts \frac{1}{4}} m (s-1) \L_m
 + {\ts \frac{1}{8}} \hat \xi ( \Lxib - \Lxi)
+ {\ts \frac{1}{128
    \alpha}} ( \Lgfb - \Lgf)\,,
\end{align}
where the square brackets denote an insertion of local composite
operators, and $\L_M, \, \L_m \,$, $\Lxi \, , \Lgf$ are defined in (\ref{L}).
The indices ${}^3_3$ stand for the UV and IR subtraction degrees.
We see that in addition to the mass terms, the matter-background
coupling and the gauge fixing terms are not Weyl invariant.
For exploring the consequences of this Ward identity further, we use
appropriate Zimmermann identities \cite{quantum, PS2}. These identities
express hard insertions in terms of the corresponding soft insertion
plus a basis of hard local field monomials with the appropriate
symmetries and quantum numbers.
Here the relevant identities are, using the basis (\ref{L}),
\begin{align}
\left[m(s-1) \L_m\right]_3^3\cdot \Gamma &= u_m m(s-1) \left[\L_m\right]_2^2\cdot\Gamma \nonumber \\ 
& \qquad + \left[u_\Phi \LPk +\ug \Lg + u_\xi \Lxi + \uxib\Lxib \right]_3^3\cdot
\Gamma \, , \label{Zim}\\[0.3cm]
\left[\hM\L_M\right]_3^3\cdot\Gamma &= v_M \hM \left[\L_M\right]_1^1\cdot\Gamma \nonumber \\
&\quad + \left[v_g\L_g + v_\xi (\Lxi + \Lxib) \right]_3^3\cdot\Gamma
\, \label{ZiM} 
\end{align}
for the mass terms as well as
\begin{align}
\left[ \Lgfb+ \Lgf  \right]_3^3 \cdot \Gamma &= \left[ \Lgfb +\Lgf
\right]_1^1  \cdot \Gamma 
+\left[ 2 r_g \L_g +2 r_\xi (\Lxi+\Lxib) \right]_3^3 \cdot \Gamma
\, , \label{Zigf1}\\[0.3cm] 
\left[ \Lgfb-\Lgf \right]_3^3 \cdot \Gamma &= \left[ \Lgfb-\Lgf \right]_1^1
\cdot \Gamma + t \left[ \Lxib-\Lxi \right]_3^3 \cdot \Gamma \quad
 \, , \label{t} \\[0.3cm]
\left[ \Igf \right]_4^4 \cdot \Gamma &= \left[ \Igf\right]_2^2 \cdot \Gamma 
+\left[ r_g I_g +r_\xi (\Ixi+\bar\Ixi) \right]_4^4\cdot \Gamma \, \label{ZiI}
\hspace{1.9cm}
\end{align}
for the gauge fixing terms.
The coefficients $u,v,r,t$ are power series in the couplings and in
$\hbar$. By integrating the identity (\ref{t}) over chiral superspace,
it may be shown that $t=0$.

In order to eliminate the symmetry breaking terms $\Lxi$ of
matter-background coupling type from the Ward identity (\ref{wweyl}),
we impose $R$ invariance
\begin{gather}
W^R \Gamma = 0 \Big|_{s=1} \, .\label{WR}
\end{gather}
Here the $R$ symmetry Ward operator is obtained from the
superconformal Ward operator combining diffeomorphisms and
Weyl transformations, by using
\begin{gather}
\Omega^{\al \, (R)} = -i \theta^\al \bar \theta^2 r \, ,  \quad
\si^{(R)} 
= {\ts
  \frac{2}{3}} i r \, , \label{or}
\end{gather}
for the parameter $\Omega^\al$ of local superconformal transformations
(\ref{sigmaomega}),
as appropriate for $R$ symmetry transformations. This gives
\begin{align}
W^R{} \Gamma &= \int\dV (\Omega^{\al \, (R)} w_\al - \bar \Omega_\da
{}^{(R)} \bar w^\da) \Gamma\,, \label{rsc} \\
w_\al &\equiv w^{(\Lambda)}_\al(H,J,\Phi,A_\pm) + {\ts\frac{1}{12}} D_\al
w^{(\si)}(A_\pm,J)\, ,
\end{align}
using the Ward operators of appendix \ref{localWardoperators}.

The imposition of $R$ invariance implies
\begin{gather} \label{xiu}
\hat \xi = u_\xi - u'{}_\xi \, ,
\end{gather}
which fixes the coupling $\hat \xi$ uniquely as a function of the
coupling $g$ and thus amounts to a reduction of couplings \cite{Reduction}.

The Ward identity (\ref{wweyl}) leads to a Callan-Symanzik (CS) equation
since from dimensional analysis we have
\begin{gather}
\mu \pr_\mu \Gamma_{\rm eff} \, = \, -i W^D \Geff \, , \qquad W^D \, =
\,
\int\dS w^{(\si)} \, + \, \int\dSb \bar w^{(\bar \si)} \, ,
\end{gather}
where $\mu$ stands for all mass parameters of the theory. 
Thus using the Zimmermann identities (\ref{Zim}) and (\ref{ZiM}) 
we obtain the CS equation
\begin{align} 
 \C \Gamma
&= \quar (1-2\gamma_A) m(s-1) [I_m+\bar I_m]_3^3\cdot\Gamma \nonumber \\
&\quad + \left(\tfr{1}{8}(1-\be)\hM + \tfr{1}{16}\be (g\pr_g
  -4\al\pr_\al)\hM \right) [I_M]_2^2\cdot\Gamma \, ,
\label{CSglobal}\\ \C &
\equiv
\mu \pr_\mu + \be \left(g \pr_g -\N_\hP-2\al\pr_\al \right) -
  \gamma_A \N_A  \label{C} \, ,
\end{align}
which holds subject to the coefficients satisfying
\begin{align}\left( 1-\be-2\be\al\tfr{\pr_\al\hM}{\hM} + \half \be g
    \tfr{\pr_g\hM}{\hM} \right) v_g + 4 (1-2\gamma_A)u_g + (\half\be g\pr_g
    - \be\al \pr_\al -\gamma_A) z_g &= 0 \, , \label{cond1} \\
\tfr{1}{64} (\be + \be\al\pr_\al -\half \be g\pr_g) z_\Phi + \half
(1-2\gamma_A) u_\Phi &=0 \, . \label{cond2}
\end{align}
These relations define the functions
$\beta$ and $\gamma_A$, which are the usual $\beta$ and
$\gamma$ functions, noting that
\begin{gather}
\beta^g = \beta \cdot g \, , \quad \beta\equiv \gamma_\Phi \, .
\end{gather}
$\N_\Phi$ and $\N_A$, the operators counting fields in $\C$, 
are given by
\begin{gather}
\N_\Phi = \int\dS w^\is(\hP) + c.c. \, , \qquad
\N_A   = \sum\limits_{\pm} \, \int\dS A_\pm \frac{\de}{\de A_\pm}
\, + \, \int \dSb \bar A_\pm \frac{\de}{\de \bar A_\pm} \, ,
\label{count}
\end{gather}
where $w^\is(\hP)$ is a power series in $H^{\al\da}$ as given in appendix
\ref{localWardoperators}. To lowest order it is given by
\begin{equation}
w^{(\si)}(\hat \Phi) \Gamma
=
\bar D^2 \left( \hat \Phi \frac{\de\Gamma}{\de \hP} \right)
\,. \label{wsigmahP}
\end{equation}
Classically, $\hP$ is Weyl invariant. However in the quantised theory it
acquires an anomalous Weyl weight $\be$. 

For the proof of the CS equation (\ref{CSglobal}) to all orders we make use
of the consistency condition
\begin{gather} \label{consistency}
[ W^R , \C ] = 0
\end{gather}
for the operators defined in (\ref{rsc}) and (\ref{C}), proceding by
induction in orders of $\hbar$. In particular for the
matter-background coupling $\Ixi+ \bar I_\xi$, (\ref{consistency}) ensures
that subject to (\ref{xiu})
\begin{gather} \label{rxi}
2(1-\beta) v_\xi + 4 (1-2\gamma_A) (u_\xi+ u'_\xi) + 2 (\beta g\pr_g
-2\gamma_A)\hat \xi + \beta g {\ts\frac{ \pr_g \hM}{\hM}} v_\xi - 4 \beta \al
\pr_\al \hat \xi = 0 \, ,  
\end{gather}
such that there are no $\Ixi+\bar I_\xi$ contributions to the r.h.s.~of the
CS equation (\ref{CSglobal}).

\subsection{Gauge parameter dependence}

The CS operator $\C$ given by (\ref{C}) involves $\pr_\al$, the
derivative with respect to the gauge parameter $\al$. 
When acting on the vertex functional $\Gamma$ this derivative yields
\begin{align}
\pr_\al \Gamma &= \,[\pr_\al \Geff ]^4_4 \cdot \Gamma \nonumber\\
&= \, [ - {\ts \frac{1}{128}} \pr_\al z_\Phi I_\Phi  + {\ts
    \frac{1}{16}} \pr_\al \hM I_M + {\ts \frac{1}{128 \al^2}} 
\Igf  + {\ts \frac{1}{16}} \pr_\al z_g I_g 
+ {\ts \frac{1}{8}} \pr_\al \hat \xi ( I_\xi + \bar I_{\xi} ) ]^4_4 \cdot
\Gamma \, . \label{al1}
\end{align}
This hard symmetry breaking term may be simplified using the
Zimmermann identity (\ref{ZiI}). Moreover we are free to choose the
normalisation conditions
\begin{gather}
\pr_\al z_\Phi=0 \, \quad \pr_\al \hM = 0 \, , \quad 8 \pr_\al z_g +
{\ts \frac{1}{\al^2}}r_g = 0 \, , \label{norma}
\end{gather}
such that (\ref{al1}) reduces to
\begin{align}
\pr_\al \Gamma &= {\ts \frac{1}{128 \al^2}} [\Igf]^2_2 \cdot \Gamma 
\, + ({\ts \frac{1}{8}} \pr_\al \hat \xi - {\ts
  \frac{1}{128 \al^2}}r_\xi) [ \Ixi + \bar I_\xi ]^4_4 \cdot \Gamma
\, .
\end{align}
The coupling $\hat \xi$ - and thus its $\al$ dependence - 
have already been fixed by imposing $R$ invariance
in (\ref{WR}). By making use of the consistency condition
\begin{gather}
[ W^R, \pr_\al] = 0
\end{gather} 
we may show that
\begin{gather}
{\ts \frac{1}{8}} \pr_\al \hat \xi + {\ts
  \frac{1}{128 \al^2}}r_\xi = 0 \, , \label{xicond}
\end{gather}
such that 
\begin{align}
\pr_\al \Gamma &= {\ts \frac{1}{128 \al^2}} [\Igf]^2_2 \cdot \Gamma 
\, . \label{al2}
\end{align}
When reducing to flat space, the gauge fixing term on the
r.h.s.~vanishes when acting on states of the physical Hilbert space,
as was shown in \cite{PS1,PS2}. This is essentially due to its
reduced subtraction degree. Thus the physical degrees of freedom 
in SQED on flat space are gauge parameter independent.
Here it is straightforward to extend this argument to Green functions
involving an insertion of the supercurrent by making use of (\ref{H})
and varying (\ref{al2}) with respect to the superpotential $H_{\al
\da}$, which gives, when restricting to flat superspace,
\begin{align}
\pr_\al [V_{\al \da}]^3_3 \cdot \Gamma \Big|_{H=0}
&=  \left[ 
V^g_{\al \da}\right]^0_0 \cdot \Gamma + {\ts \frac{1}{128 \alpha}} [V_{\al 
\da}]^3_3 \cdot 
[ \Igf ]^2_2 \cdot \Gamma \Big|_{H=0} \, , \\
V^g_{\al \da} &\equiv {\ts \frac{1}{16 \al} } \, 
\frac{\de \Igf}{\de H^{\al \da}} \, , \label{Vg}
\end{align}
with $V^g{}_{\! \al \da}$ the gauge non-invariant part of the supercurrent.
Each of the terms on the r.h.s.~have been shown to be unphysical on
flat space, as discussed in \cite{PS2}. Thus when acting on states of
the physical Hilbert space, the supercurrent insertion is gauge
parameter independent. 

\subsection{Local Callan-Symanzik equation \label{lcs}}

Just as for the Wess-Zumino model \cite{ERS1}, we derive a
{\it local} CS equation, i.e.~we express the anomalies present in the
local chiral Weyl symmetry Ward identity (\ref{wweyl}) in such a way
that upon integration and combination with the corresponding
antichiral equation we recover the {\it global} CS equation
(\ref{CSglobal}). For this purpose we introduce the chiral effective
Lagrangian
\begin{align}
\Leff \equiv& \, - {\ts \frac{1}{256}} z_\Phi  \L_\Phi  + {\ts \frac{1}{32}}
 z_g \L_g + \quar m (s-1) \L_m + {\ts \frac{1}{32}} \hM \L_M
 + {\ts \frac{1}{8}} (\hat \xi - \hat \eps) \L_\xi + {\ts \frac{1}{8}}
 \hat \eps  \Lxib \nonumber\\ & 
- {\ts \frac{1}{512 \alpha}} (\Lgf +
 \Lgfb) \,,
\end{align}
using the basis (\ref{L}), such that
\begin{gather}
\Geff = \int\dS \Leff + \int\dSb \Leffb \, 
\end{gather}
with $\Geff$ as given by (\ref{Geff}).
$\hat \eps$ is an additional coupling parameter which does not
contribute to $\Geff$ due to
\begin{gather}
\int\dS (\Lxi - \Lxib) + c.c. = 0 \, . 
\end{gather}
Then from the Ward identity (\ref{wweyl}) and the relations
(\ref{cond1}), (\ref{cond2}) and (\ref{rxi}) we obtain the local CS equation
\begin{align}
 w^{(\si,\ga)} \Gamma&=
-\be g \left[ \pr_g  \Leff \right]^3_3 \cdot \Gamma \nonumber
\\[.75ex]
& \quad - \tfr{1}{128\al} \left[ \Lgfb-\Lgf\right]^1_1\cdot\Gamma 
+ \tfr{1}{256\al} \be \left[ \Lgf+\Lgfb \right]_1^1 \cdot \Gamma \nonumber
\\[.75ex]
& \quad +  \tfr{1}{16} v_M (1-\be +\half \be g \pr_g) \hM
 \left[ \L_M\right]^1_1 \cdot \Gamma \nonumber \\[.75ex]
&\quad + \quar (1-2\gamma_A) u_m m(s-1) [\L_m]_2^2 \cdot \Gamma \, , \label{CSlocal}
\\[1ex]
w^{(\si,\ga)} \Gamma
&\equiv \left( w^{(\si)}\Gamma - 
\gamma_A A_\pm \frac{\de\Gamma}{\de A_\pm} - \half \beta
w^\is(\hat \Phi) \Gamma \right) \,
, \label {wga}
\end{align}
which holds subject to the additional condition
\begin{align}
\beta g \pr_g \hat \eps -\half \be g \pr_g \hat\xi + 3 \gamma_A \hat \xi  &=0 \,  \label{a}
 \end{align}
for the coupling $\hat \eps$. In (\ref{wga}), $w^\is(\hP)$ is given by
(\ref{count}) and (\ref{wsigmahP}).
The local CS equation (\ref{CSlocal}), which is chiral, characterises all
dynamical anomalies of 
superconformal symmetry. The global CS equation
(\ref{CSglobal}) is recovered by integrating (\ref{CSlocal}) and adding the
complex conjugate, where, as far as the gauge fixing terms are concerned,
the agreement is due to (\ref{al2}).
 From the local CS equation (\ref{CSlocal}) we see that the breakdown of
superconformal invariance manifests itself in anomalous dimensions for
the fields and in an insertion of $\Leff$ with the usual $\beta$
function as coefficient. Moreover the mass terms break conformal
symmetry as expected. In (\ref{CSlocal}) we may now set $s=1$. 
Furthermore there are gauge fixing terms breaking superconformal
symmetry. However these do not contribute when acting on states of the
physical Hilbert space, as is discussed in section \ref{gaugeinvariance} below.

\subsection{Transformation properties of the supercurrent}

By combining the local Callan-Symanzik equation (\ref{CSlocal}) with
the local Ward identity expressing diffeomorphism invariance, we
obtain the superconformal transformation {proper\-ties} of the
supercurrent. In agreement with the discussion of section 2 
and using the local Ward operators listed in appendix
\ref{localWardoperators} and in (\ref{wga}), we have,  restricting the transformation parameters
$\Omega^\al$ and $\sigma$
to be of the superconformal form (\ref{sigmaomega}), 
\begin{gather} \label{scwi}
\int\dV \Omega^\al \left( w^{(\Lambda)}{}_{\! \al} + {\ts
    \frac{1}{12}}  D_\al
w^{(\si,\ga)}  \right) \Gamma +c.c. = \int\dS \si [S^{(\ga)}]\cdot \Gamma +
    c.c. \, ,
\end{gather}
where $S^{(\ga)}$ stands for the r.h.s.~of the local CS equation (\ref{CSlocal}).
 From this superconformal Ward identity we obtain the superconformal
transformation properties of the supercurrent insertion by varying
(\ref{scwi}) with respect to $H_{\al \da}$ and subsequently
restricting to flat superspace, which yields
\begin{gather}
\W{A\hP}{\Lambda\sigma}^{(\gamma_A,\be)}(\Omega_{\rm conf})
[V_{\al\da}(z)]\cdot\Gamma \hspace{11cm} \nonumber
\end{gather} \vspace{-1cm}
\begin{align}
 &= \left[ \delta V_{\al\da}(z) \right]
\cdot \Gamma \nonumber\\
&\quad +\int\dS' \sigma(z') \left\{ -\beta g\pr_g \Leff(z') \, \cdot V_{\al\da}(z)
\right\} \cdot \Gamma + c.c. \nonumber \\
&\quad -\frac{1}{16\al} \fdq{}{H^{\al\da}(z)} \left[ \int \dS' \sigma(z')
\left(  (\Lgfb-\Lgf) - \half \beta (\Lgf+\Lgfb) \right)
 + c.c. \right]_2^2 \cdot \Gamma \nonumber \\
&\quad + \!\tfr{1}{16}v_M \left(1\!-\!\be(1\!+\!\half g\pr_g)\right) \hM
 \fdq{}{H^{\al\da}(z)} \left[ \int \dS\!' \sigma(z') \L_M(z') + c.c. \right]_2^2
 \cdot \Gamma \nonumber \\
& \quad -\half\be \left( D_\al \sigma \bar D_\da \hP -\bar D_\da \bar\sigma
  D_\al \hP \right) \fdq{\Gamma}{\hP} \, , \label{scv}
\end{align}
where we have set $s=1$. Here the flat space superconformal Ward
operator is given  by
\begin{align}
\W{A\hP}{\Lambda\sigma}^{(\gamma_A,\be)}(\Omega_{\rm conf}) &= \int\dV
\Omega^\al_{\rm conf} \left( w^{(\gamma_A)}{}_{\!\al} (A_\pm) +
w^{(\beta)}{}_{\!\al}(\hat \Phi)  \right) \, + c.c. \, , \\ 
w^{(\gamma_A)}{}_{\!\al}(A_\pm) &= \quar D_\al A_\pm \frac{\de}{\de A_\pm} - {\ts
  \frac{1}{12}} (1+\gamma_A) D_\al \left( A_\pm \frac{\de}{\de A_\pm}\right) \, ,
\nonumber\\  
w^{(\beta)}{}_{\!\al}(\hP) &= 
\half \bar D^\da \left(  \bar D_\da D_\al \hP \frac{\de}{\de \hP}\right)
+ \quar \bar D^2 \left( D_\al \hP \frac{\de}{\de \hP} \right) 
- {\ts \frac{1}{12}}\beta D_\al\bar D^2 \left(\hP\frac{\de}{\de \hP} \right)
\, , \nonumber
\end{align}
where the superconformal transformation parameter $\Omega_{\rm conf}$
satisfies the condition (\ref{sigmaomega})  and where
$w^{(\gamma_A)}{}_{\!\al}(A_\pm) $
and $ w^{(\beta)}{}_{\!\al}(\hat\Phi)$ are flat space superconformal Ward
operators. These are  obtained from the flat space restriction of the
combined diffeomorphism and Weyl operators. Moreover the classical
superconformal transformation of the supercurrent is given by
 \begin{align} 
\label{Vtransclass}
\de V_{\al \da}(z) &= (\Lambda - {\ts \frac{3}{2}} (\si
   + \bar \si)) V_{\al \da}(z) \,  \\ &=
\left\{ D_\beta, \bar D_\db \right\}
\big( \bar D^\db \Omega^\beta V_{\al \da} \big) + \big(\left\{ D_\al,
\bar D_\da \right\} \bar D^\db \Omega^\beta \big) V_{\beta \db} + D^
\beta \left( \bar D^2 \Omega_\beta V_{\al \da} \right) + c.c. \, . \nonumber
\end{align}
For the classical theory without gauge fixing, the supercurrent $V_{\al \da}$
is  conformally covariant, i.e.~a quasi-primary field.
Furthermore in (\ref{scv}), for any insertion $[S]$  the curly brackets denote
\begin{gather}
\{ V_{\al \da}(z)\cdot S(z') \} \cdot \Gamma \equiv 8 \frac{\de}{\de H^{\al
    \da}(z)} [S(z')] \cdot \Gamma = [V_{\al \da}(z)] \cdot [ S(z')] \cdot
    \Gamma + 8 \left[ \frac{\de S(z')}{\de H^{\al\da}(z)} \right]
    \cdot \Gamma \, .
\end{gather}
Finally the last term on the r.h.s.~of (\ref{scv}) is due to the $H$
dependence of $\hP$ as given by (\ref{gaugeflat}), which leads to an anomalous
conformal transformation of $\hP$ in the quantised theory. This
reflects the fact that after quantisation, the gauge field $\hP$ is no 
longer quasi-primary.

\section{Gauge invariance \label{gaugeinvariance}}
\setcounter{equation}{0}

The superconformal transformation of the supercurrent as given by
(\ref{scv}) contains terms involving the gauge fixing and thus is
seemingly not gauge invariant. However this is only apparent. In this
section we show that when acting on elements $\phys$
of the Hilbert space of physical states, with infinitesimal
superconformal transformation $\dphys$, then
\begin{gather}
\Delta \left( \lphys 
V_{\al \da} \phys \right) = \lphys \left( 
\Delta V_{\al  \da} \right) \phys + \ldphys V_{\al \da} \phys+
 \lphys V_{\al \da} \dphys  \label{phystransform}
\end{gather}
is gauge invariant. $\Delta$ is the superconformal transformation 
corresponding to $\W{A\hP}{\Lambda\sigma}^{(\gamma_A,\be)}(\Omega)$ acting on $[V_{\al\da}(z)]\cdot\Gamma$ in (\ref{scv}).
For showing gauge invariance we note that the physical states satisfy
\begin{gather} \label{pp}
(D^2 \Phi)^{(-)} \phys = (\bar D^2\Phi)^{(-)} \phys = 0 \, , \quad
\lphys (D^2 \Phi)^{(+)}  = \lphys (\bar D^2\Phi)^{(+)} = 0 \, ,
\end{gather}
where $(-), (+)$ stand for contributions involving annihilation
or creation operators, respectively.
Requiring the conditions (\ref{pp}) to be invariant under an anomalous
Weyl transformation with weight $\beta$, in agreement with (\ref{wga}), we have
\begin{align}
0 &= \de ( \lphys \bar D^2 \Phi \phys) = \, (1-\half\beta) \lphys 
\bar D^2(\bar \si \Phi)
\phys + \ldphys \bar D^2 \Phi \phys + \lphys \bar D^2 \Phi \dphys \, ,
 \nonumber\\
0 &= \de ( \lphys  D^2 \Phi \phys) = \, (1-\half\beta) \lphys \bar D^2( \si\Phi)
\phys +  \ldphys D^2 \Phi \phys +  \lphys D^2 \Phi \dphys \, . \label{pp2} 
\end{align}
Furthermore using (\ref{pp}) we have
\begin{gather} \label{pp3}
\lphys \bar D^2 (\bar \si \Phi) \phys = \lphys 
(2 \bar D_\da \bar \si \bar D^\da
\Phi) \phys \, .
\end{gather}
Using these relations we find
for the gauge non-invariant 
part $V^g{}_{\! \al \da}$ of the supercurrent defined in
(\ref{Vg}), when acting on the transformation of the physical states
$\dphys$,
\begin{align}
\ldphys V^g{}_{\! \al \da} \phys+\lphys V^g{}_{\! \al \da} \dphys
& = (1-\half\beta) \, \lphys 
\left( T_{\al \da} + c.c.\right) \, \phys \, 
 \, , \label{Vgdphys} \end{align} with
\begin{align}
T_{\al \da} &\equiv \tfr{1}{24\al} \left( [D_\al,\bar D_\da] \Phi
  \bar D^2(D^\beta \si D_\be \Phi)  
- \Phi \bar D_\da D_\al \bar
  D^2(D^\beta \si D_\beta \Phi) + \bar D_\da\Phi D_\al \bar
  D^2(D^\beta\si D_\beta \Phi)\right) \, . \label{K}
\end{align} 
Furthermore by explicit calculation we find for the contribution 
of the gauge fixing terms to $\lphys \Delta V_{\al \da}
\phys$ in (\ref{phystransform}) as given by (\ref{scv})
\begin{align}
 \lphys \Delta V_{\al \da} |_{\rm gf}
\phys &\equiv
-{\ts\frac{1}{16\al}} \lphys  \fdq{}{H^{\al\da}(z)} \Big(
 \int \dS' \sigma(z')
\Big(  (\Lgfb-\Lgf) \nonumber\\
& \hspace{7cm} - \half \beta (\Lgf+\Lgfb) \Big) +c.c. \Big) \phys
\nonumber\\ 
& = - (1-\half\be) \lphys (T_{\al \da} + c.c.) \phys \, ,
 \label{gfphys} 
\end{align}
with $T_{\al \da}$ as in (\ref{K}). Therefore the
contributions of (\ref{Vgdphys}) and (\ref{gfphys}) to
(\ref{phystransform}) cancel each other, 
\begin{gather} 
\lphys \Delta V_{\al \da} |_{\rm gf}
\phys 
+ \ldphys V^g{}_{\! \al \da} \phys+\lphys V^g{}_{\! \al \da} \dphys=0
 \, . 
\end{gather}
Thus the conformal
transformation of the gauge fixing contribution to the supercurrent is
compensated by the conformal transformation of the physical states. 
Furthermore we show that the remaining contributions to (\ref{scv})
vanish between physical states.
Using the results of \cite{PS2} we have 
\begin{gather}
\lphys \de V^g {}_{\!\al \da} \phys \equiv \lphys 
\left(\left(\Lambda - {\ts \frac{3}{2}}(
\si + \bar \si) \right) V^g{}_{\!\al \da}\right) \phys = 0 \,  
\end{gather}
for the contribution of the gauge fixing term to the classical
superconformal transformation of the supercurrent.
Furthermore the gauge field contact terms in (\ref{scv}) may be
shown to be gauge invariant by making use of the Zimmermann
identity (\ref{Zigf1}) once more. Since for superconformal
transformations $\pr_a \si = \pr_a \bar \si$, we have for the last term in
(\ref{scv}) 
\begin{gather}
\left( D_\al \si \bar D_\da \Phi - \bar D_\da \bar
  \si D_\al \Phi \right) \frac{\de \Gamma}{\de \Phi} = 
\bar D_\da \left( D_\al \si \,  \Phi\frac{\de
  \Gamma}{\de \Phi}\right) 
 - D_\al \left( \bar D_\da \bar
  \si \,  \Phi  \frac{\de \Gamma}{\de \Phi}\right) \, .
\end{gather}
The gauge fixing terms contributing to this expression are given by
\begin{gather} \label{bgf}
\Phi \frac{\de \Gamma}{\de \Phi} \Big|_{\rm gf} = \, - \, {\ts
  \frac{1}{128 \al} }\left[ \Phi D^2 \bar D^2 \Phi + \Phi \bar D^2 D^2
  \Phi\right]^2_2  \cdot \Gamma \, .
\end{gather}
Without chiral projection, the Zimmermann identity (\ref{Zigf1}) 
reads
\begin{align}
\left[ \Phi D^2 \bar D^2 \Phi + \Phi \bar D^2 D^2
  \Phi\right]^2_2  \cdot \Gamma &= \left[ \Phi D^2 \bar D^2 \Phi + 
\Phi \bar D^2 D^2
  \Phi\right]^0_0  \cdot \Gamma \label{Zivector}\\ & 
+ \left[ r_g (\bar A_+ \e^{g
  \Phi} A_+ + \bar A_- \e^{-g \Phi} A_- )  + r_\xi (A_+ A_- + \bar A_+
  \bar A_- ) \right]^2_2 \cdot \Gamma \nonumber
\end{align}
on flat space.
Thus the gauge fixing terms contributing to (\ref{bgf}) are not
subtracted and therefore vanish between physical states
by virtue of (\ref{pp}). The remaining terms in (\ref{Zivector}) are
gauge invariant. - This completes our proof that (\ref{phystransform})
with $\Delta V_{\al \da}$ given by (\ref{scv}) is gauge invariant for
physical states.

Similarly we may show that the local CS equation (\ref{CSlocal}) is
gauge invariant when sandwiched between physical states. For this we generalise
(\ref{pp}) to curved superspace. The operators $D^2 \Phi$, $ \bar D^2
\Phi$ are solutions of the flat space chiral equations of motion
\begin{gather} \label{flatem}
{\ts \frac{1}{16}} \bar D^2 D^2 A - m^2 A =0 \, , \quad
{\ts \frac{1}{16}}  D^2 \bar D^2 \bar A - m^2 \bar A =0 \, ,  
\end{gather}  
with $A \equiv \bar D^2 \Phi$, $\bar A \equiv  D^2 \Phi$, such that
$D^2 \Phi$, $\bar D^2 \Phi$ are free fields.  
The curved space generalisation of (\ref{flatem}) is obtained by
varying the action
\begin{align}
\Gamma_2 &= \, -\, {\ts \frac{1}{128\al}} \int\dV E^{-1} \Phi \Pc
  \Pac  \Phi + {\ts \frac{1}{16}}
M^2 \int\dV E^{-1} \Phi^2 \nonumber\\ & \, - {\ts \frac{1}{64 \al}}
\int\dV E^{-1} \left( R \Phi \Pc  \Phi +\bar R
  \Phi  \Pac
   \Phi \right) \label{Gamma2}
\end{align}
given in the chiral representation.
This action consists of the gauge fixing and mass terms as well as of
an additional term coupling the gauge field to the supergravity background.
For
\begin{gather} \label{ff}
A \equiv \Pc \Phi \, , \quad \bar A \equiv \Pac \Phi
\end{gather}
we obtain 
\begin{gather}
{\ts \frac{1}{16}} \Pc \left( \Pc  \bar A +  \Pc
   A \right) \hspace{6cm}\nonumber\\ \hspace{5cm}
+ \quar \Pc (RA+ \bar R \bar A) - m^2 A = 0 \, ,
  \nonumber\\  
{\ts \frac{1}{16}} \Pac \left( \Pac   A +  \Pac
  \bar A \right) \hspace{6cm}
\nonumber\\ \hspace{5.9cm}+ \quar \Pac (RA+ \bar R \bar A) - m^2 \bar A
  = 0\, ,
\end{gather}
with $m= \sqrt \alpha M$, 
which is the equation of motion for chiral fields with
minimal coupling to supergravity, $\xi=\quar$ for $\xi$ as in
(\ref{Geff}). Thus (\ref{ff}) is a solution of the free chiral
equation of motion on curved superspace, such that the physical
Hilbert space is given by 
\begin{align} \label{ppc}
(\Pc \Phi)^{(-)} \phys &= (\Pac\Phi)^{(-)} \phys = 0 \, , \\
\lphys (\Pc \Phi)^{(+)}  &= \lphys (\Pac \Phi)^{(+)} = 0 \, , \nonumber
\end{align}
in analogy to (\ref{pp}). Using this it is immediately obvious that
the local CS equation (\ref{CSlocal}) is gauge invariant 
between physical states. - The last term in (\ref{Gamma2})
has not been taken into account in the discussion of the local CS
equation in section \ref{lcs}. However using (\ref{ppc}) 
these terms may be easily seen
to vanish between physical states. 

\section{Conclusion}
\setcounter{equation}{0}

 From (\ref{scv}) it is straightforward to obtain the superconformal
 transformation properties for connected Green functions with an insertion
 of the supercurrent. We find 
\begin{equation}
\delta \l V_{\al\da}(z) \, X \r = \int \dS' \sigma(z') \l \be g\pr_g
\Leff(z') \cdot V_{\al\da}(z) \, X \r + c.c. + \text{soft terms}\,,
\label{Grftrans}
\end{equation}
where 
\begin{equation}
X=A_\pm(z_1)\dots A_\pm(z_n)\bar A_\pm(z_{n+1})\dots \bar
A_\pm(z_m)\,.
\label{Xdef}
\end{equation}
Since the connected Green functions are defined according to the
Gell-Mann-Low formula, the discussion of section 4 applies and
(\ref{Grftrans}) is gauge invariant. The superconformal transformation
$\delta$ is defined by
\begin{align}
\delta \, \l V_{\al\da}(z) \, X \r = & \phantom{+} \l \delta V_{\al\da}(z) \,
X \r \nonumber \\
& + \sum_{k=1}^n \l V_{\al\da}(z) \, A_\pm(z_1) \dots \delta A(z_k) \dots
  A_\pm(z_n) \, \bar A_\pm(z_1') \dots \bar A_\pm(z_m')\r \nonumber \\
& + \sum_{k=1}^m \l V_{\al\da}(z) \, A_\pm(z_1) \dots
  A_\pm(z_n) \, \bar A_\pm(z_1') \dots \delta \bar A_\pm(z_k') \dots \bar
  A_\pm(z_m')\r \,,  
\label{deltadef}
\end{align}
with $\delta V_{\al\da}(z)$ the classical conformal transformation 
given by (\ref{Vtransclass}), and 
\begin{equation}
\delta A_\pm = \left( \Lambda -(1+\gamma_A) \, \sigma \right) \, A_\pm \,, \quad
\delta \bar A_\pm =  \left( \bar \Lambda -(1+\gamma_A) \, \bar \sigma \right) \,
\bar A_\pm \,.
\end{equation}
We consider matter fields only in (\ref{Xdef}) for simplicity. For gauge
fields $\hP$ there are additional terms arising from the last term in
(\ref{scv}). 

In this paper we have determined  a suitable field structure for
discussing both conformal transformations and  gauge
invariance, from which we have obtained Ward operators which ensure
conformal covariance of the supercurrent. 
Our result (\ref{Grftrans}) states that there is a gauge invariant
expression for the superconformal transformation properties of Green
functions with a supercurrent insertion which is well-defined to all orders
in perturbation theory, including the
soft mass breaking terms. Discarding these
we see that $\l V_{\al\da} \cdot  X\r$ is conformally invariant
when the $\beta$ function vanishes (which for SQED considered here applies
just to the free theory). For superconformal theories as relevant to RG
fixed points, correlation functions involving the supercurrent
have been constructed
explicitly in \cite{osborn} using the symmetry constraints.

So far we have considered one insertion of the supercurrent only. 
Ward identities involving Green functions with multiple insertions 
of the supercurrent, $\l V_{\al \da}(z_1) \cdot V_{\be \db}(z_2) \cdot
\dots \cdot V_{\omega \dot \omega}(z_n) \cdot X \r$, may in principle be
obtained by varying (\ref{scv}) an appropriate number of times with
respect to the supergravity prepotential $H$. However in this case the
proof of gauge invariance in analogy to the calculation performed in
section 4 gets very tedious since higher orders in $H$ have to be
considered explicitly. In this case a BRS approach seems more
appropriate, which applies of course also to a possible generalisation
to non-abelian gauge theories\footnote{For a complementary approach
  using the component formalism see \cite{PW}.}.

As a final remark we discuss the consequences of $R$ invariance.
We write the $R$ symmetry operator defined in (\ref{rsc}) in the form 
\begin{equation}
W^R \Gamma \equiv i \int \dx w\Gamma\,. \label{wR}
\end{equation}
For the subsequent discussion, it is important to remember that $w$
contains both 
diffeomorphisms and Weyl transformations.
$R$ invariance implies
\begin{align}
W^R \Gamma =0 \;\;  &\Rightarrow \;\; 
w\Gamma = {\rm total \; divergence} \, .
\end{align}
 Calculating 
$w\Gamma$ as given by (\ref{wR}) 
we find, when restricting to flat space,
\begin{equation}
\l \pr^a V_a \cdot X \r = i \l (D^2 S - \bar D^2 \bar S) \cdot X \r
+ \sum_j f^R(z-z_j) \l X'\r\,, \label{VS}
\end{equation}
where $S$ is the Weyl symmetry breaking term given by (\ref{wweyl}).
Evaluating this term with the help of the Zimmermann identities
(\ref{Zim}), (\ref{ZiM}) we find  
\begin{align}
i \l (D^2 S - \bar D^2 \bar S) \cdot X \r &= 8 i \, \pr^a \left( 2 \pr_a +
  i (D \sigma \bar D)_a\right) \l B \cdot X\r \,, \\[1ex]
B &= B_5 + B_\Phi \,, \label{Bdef}\\
B_5 &= - \tfr{1}{24} (4u_g + v_g) \left( \bar
  A_+\e^{g\hP}A_+ + \bar A_- \e^{-g\hP}A_- \right) \\
& \quad - \tfr{1}{24} (2u_\xi + 2u_\xi' + v_\xi)\left( \bar A_+ \bar A_- +
  A_+ A_- \right)  -\tfr{1}{24} \hM \hP^2 \,,  \nonumber\\
B_\Phi &= -\tfr{1}{6} u_\Phi \left( D^\al  \hP \bar D^2 D_\al  \hP + \bar
  D_\da  \hP D^2
  \bar D^\da  \hP + \hP D^\al  \bar D^2 D_\al  \hP \right) \,. \label{B5def}
\end{align}
$V_a$ is the supercurrent of SQED as given by (\ref{H}) with the
action (\ref{Geff}). $B_\Phi$ arises from the gauge field term $D^2 F^\al
F_\al - \bar D^2 \bar F_\da \bar F^\da$.
The Zimmermann coefficients $u_g, u_\xi, u_\xi', u_\Phi$, $v_g$, $v_\xi$ are
related to the $\beta$ and $\gamma$ functions by virtue of
(\ref{cond1}), (\ref{cond2}) and (\ref{rxi}).
 The terms involving $f^R(z-z_j)$ in (\ref{VS}) 
are contact
terms arising from varying $w\Gamma$ as given by (\ref{wR})
with respect to the matter fields
in order to obtain $X$ as in (\ref{Xdef}).
It is a specific feature of the Weyl anomalies present in SQED with
massless matter fields that
they may be written  as total divergences in the $R$ symmetry Ward
identity.
This is due to the fact that they are of the form $S=\bar D^2 B$.

As follows from the discussion of \cite{PS2}, the supercurrent $V_a$ in
(\ref{VS}) as given by
(\ref{H}) has the component decomposition 
\begin{align}
V_a (z) =& \, R_a(x) - i \theta^\al ( Q_{a \, \al} (x) - {\ts \frac{1}{3}}
 (\si_a\bar
\si_b )_{\al}{}^{\be}Q^{b}{}_\be ) + i 
\bar \theta_\da ( \bar Q_a{}^\da (x) - {\ts \frac{1}{3}} (\bar \sigma_a
 \sigma_b)^\da {}_\db \bar Q^{b\db})
\nonumber \\ &  - 4 (\theta \si \bar \theta)^b
( T_{ab}(x) - {\ts \frac{1}{3}} \eta_{ab} T_c{}^c(x) + \quar 
\eps_{abcd} \pr^c R^d) + \cdots \, ,
\end{align}
with the $R$ current, the supersymmetry currents $Q_{a\al}$, $\bar
Q_{a\da}$ and the energy momentum tensor $T_{ab}$.
Furthermore we may define a current 
\begin{equation}
V_{\al\da}^B = V_{\al\da} + 4 [D_\al, \bar D_\da] B\,, \label{VB}
\end{equation}
with $B$ as in (\ref{Bdef}), whose divergence is anomaly-free, 
\begin{equation}
\l \pr^a V^B_a \cdot X \r = \sum_j f^R(z-z_j) \l X'\r\,.
\end{equation}
This is referred to as changing from so-called `S' to `B' breaking in
\cite{PS2} (see also \cite{ERS1}).  
The current (\ref{VB})
has the component decomposition
\begin{gather}
V^B_a (z) = R_a(x) + \theta^\al Q_{\al \, a}(x) + \bar \theta_\da \bar 
Q^\da{}_a (x)
+ (\theta \si \bar \theta)^b T_{ab} (x) + \dots\,.
\end{gather}
We note that $V^B_a$ is not gauge invariant since $B_\Phi$ is not gauge
invariant.

Finally we discuss a current whose divergence agrees with the chiral
anomaly. The current 
\begin{align}
V_{\al\da}' &= V_{\al\da} + K_{\al\da}  \,, \label{Vp}\\
K_{\al\da} &= 4 [D_\al, \bar D_\da] B_5 \nonumber \\
&=  - {\ts \frac{1}{3}} [D_\al, \bar D_\da] \Big(  (2 u_g
  +\half v_g)
( \bar A_+ \e^{g\hP}  A_+ + \bar A_- \e^{-g\hP} A_-) \nonumber\\ &
\hspace{5cm} + (u_\xi + u_\xi' +
  \half v_\xi)
(A_+A_- +\bar A_- \bar A_+)   + \half \hM \hP^2 \Big) \, , \label{Kcurrent}
\end{align}
which coincides with $V^B_{\al\da} - 4 [D_\al, \bar
D_\da] B_\Phi$, satisfies the Ward identity 
\begin{align}
\l \pr^a V'_a \cdot X \r  =&\, - \, {\ts
 \frac{1}{6}} \, i \, u_\Phi \,\l 
(D^2 F^\al F_\al - \bar D^2 \bar F_\da \bar F^\da) \cdot X \r
\nonumber\\
& + \sum\limits_{j} f^R(z-z_j) \l X' \r \,  . \label{AB} 
\end{align}
$V'_a$ is gauge invariant except for the soft mass term.
Its significance arises from the fact 
that the coefficient $u_\Phi$ of its anomaly has 
no corrections beyond one loop \cite{PS3}.  This same
coefficient appears in the anomaly of the so-called Konishi
current \cite{Konishi}, which is the current associated with $R$
symmetry on flat space without coupling to supergravity, where the
chiral local
$R$ symmetry Ward operator is given by
\begin{equation}
w_5 \equiv A_+ \frac{\de}{\de A_+} + A_- \frac{\de}{\de A_-} \, .
\end{equation}
Therefore the current $V'_a$ as given by (\ref{Vp}) 
satisfies an Adler-Bardeen
theorem in the sense that its gauge anomaly has no quantum corrections
beyond one loop.  It has to be noted however that the higher $\theta$
components of $V'_a$ 
neither correspond to the supersymmetry currents nor to the
energy-momentum tensor.

(\ref{AB}) shows that although the chiral anomaly and  the
superconformal trace anomaly as expressed by the divergence of the
supercurrent are related, the trace anomaly and thus the $\beta$
functions may acquire higher order corrections while the chiral
anomaly is one-loop. This issue is discussed from  
a different point of view in \cite{Arkani}. The discussion here shows how
the geometrical structure of the supergravity background relates the
two anomalies. Effectively the Konishi anomaly appears as a term
breaking Weyl invariance, 
whereas the supercurrent is present in the
diffeomorphism Ward identity since it is coupled to the prepotential
$H^{\al \da}$. Thus in the superconformal Ward
identity which relates Weyl symmetry and diffeomorphisms both currents are
present.

\appendix

\newpage

\renewcommand{\theequation}{\thesubsection.\arabic{equation}}

\section{Appendix}

\subsection{Infinitesimal Transformations \label{infinitesimal}}
\setcounter{equation}{0}
Here we list the infinitesimal symmetry transformations for the symmetries
discussed in section 2. For the supergravity prepotential $H^{\al\da}$ and
the gauge field $\hP$, these are power series in $H$.

\noindent{\bf Diffeomorphisms}
\begin{align}
 \delta_\Omega A_\pm &=  {\ts \frac{1}{4}} \bar D^2 \left(  \Omega^\alpha
 D_\alpha A_\pm  \right)
 \\[1ex] 
\delta_\Omega H^{\alpha\da} &= \half \bar D^\da \Omega^\alpha
 \nonumber\\ 
&\quad + \quar \bar D^{\dot\beta} \Omega^\beta
 \{D_\beta,\bar D_{\dot\beta} \} H^{\alpha\da} - \quar
 H^{\beta\dot\beta} \{D_\beta, \bar D_\db \} \bar D^\da \Omega^\alpha
 + \quar \bar D^2 \Omega^\beta D_\beta H^{\alpha\da}
                         \label{delta_Omega_H} \nonumber \\
& \quad + O(H^2)  \\[1ex]
\delta_\Lambda J &= \quar \bar D^2 \left( \Omega^\al D_\al J \right) +
\tfr{1}{12} \bar D^2 D^\al \Omega_\al \\[1ex]
\delta_\Lambda \hP &= \delta_{\Lambda}^{(0)} \hP + \delta_{\Lambda}^{(1)}
\hP + O(H^2) \nonumber \\[1ex]
\delta_{\Lambda}^{(0)} \hP
&= \half \bar D^\da \Omega^\al \bar D_\da D_\al \hP + \quar \bar D^2
\Omega^\al D_\al \hP + c.c. \\[1ex]
\delta_{\Lambda}^{(1)} \hP
&= \tfr{1}{8} H^{\al\da} \bar D^\db \Omega^\be \left( [D_\al, \bar D_\da]
  [D_\be, \bar D_\db] - \{ D_\al, \bar D_\da\} \{ D_\be, \bar D_\db\}
\right) \hP \nonumber \\
& \quad - \tfr{1}{8} H^{\al\da} \bar D^2 \Omega_\al D^2 \bar D_\da \hP +
\quar H^{\al\da} D_\al \bar D_\da \Omega^\be \bar D^2 D_\be \hP + c.c.
\end{align}

\noindent{\bf Weyl Transformations}
\begin{align}
\delta_\sigma A_\pm &= -\sigma A_\pm \\[1ex]
\delta_\sigma H^{\al\da} &= 0 \\[1ex]
\delta_\sigma J &= \sigma  \\[1ex]
\delta_\sigma \hP &= \delta_\sigma^{(0)}\hP + \delta_\sigma^{(1)}\hP
+O(H^2)\nonumber \\[1ex]
\delta_\sigma^{(0)}\hP &= (\sigma+\bar \sigma) \hP \\[1ex]
\delta_\sigma^{(1)}\hP &= -\half H^{\al\da} \{ D_\al, \bar D_\da \}
(\sigma-\bar \sigma) \hP +\half (\sigma+\bar \sigma) H^{\al\da} [D_\al, \bar
D_\da] \hP \nonumber \\
&\quad -\half H^{\al\da}[D_\al, \bar D_\da] \left( (\sigma+\bar\sigma) \hP \right)
\end{align}

\noindent{\bf Gauge Transformations}
\begin{align}
\delta_\lambda \hP &= i (\lambda-\bar \lambda) \\[1ex]
\delta_\lambda A_\pm &= \mp ig A_\pm
\end{align}

\subsection{Local Ward operators \label{localWardoperators}}
Here we list the local Ward operators corresponding to the infinitesimal
transformations of \ref{infinitesimal}.

\noindent{\bf Diffeomorphisms}
\begin{align}
w_\alpha\il (A_\pm)&= \quar D_\alpha A_\pm \fdq{}{A_\pm}, \\[1ex]
w_\alpha\il (J)&= \quar D_\alpha J \fdq{}{J} 
- {\textstyle \frac{1}{12}} D_\al \fdq{}{J},\\[1ex]
w_\alpha\il (H)&= {}w_\alpha^{(0)}(H)\,+\,w_\alpha^{(1)}(H)\,+\,w_\alpha^{(2)}(H)\,+\,O(H^3),\\[1ex]
w_\alpha^{(0)}(H) &= \half \bar D^\da \fdq{}{H^{\alpha\da}}, \\[1ex]
w_\alpha^{(1)}(H) &= \quar \bar D^\da \left( \{ D_\alpha,\bar D_\da \}
         H^{\beta\db} \fdq{}{H^{\beta\db}} \right)
         +\quar \{D_\beta,\bar D_\db \} \bar D^\da \left( H^{\beta\db}
         \fdq{}{H^{\alpha\da}}\right) \nonumber\\[1ex]
    &\quad  +\quar \bar D^2 \left( D_\alpha H^{\beta\db} \fdq{}{H^{\beta\db}} 
         \right),\\[1ex]
w^{(\Lambda)}_\al(\hat \Phi) &= w^{(\Lambda)}_\al(\hat \Phi)^{(0)} +
w^{(\Lambda)}_\al(\hat \Phi)^{(1)}+ O(H^2) \\[1ex]
w^{(\Lambda)}_\al(\hat \Phi)^{(0)} &= \half \bar D^\da \left( \bar D_\da
  D_\al \hat \Phi \fdq{}{\hat \Phi}\right) + \quar \bar D^2 \left(D_\al \hP
  \fdq{}{\hP} \right) \\[1ex]
w^{(\Lambda)}_\al(\hat \Phi)^{(1)} &= \tfr{1}{8} \bar D^\da \left(
  H^{\be\db} [D_\be, \bar D_\db] [D_\al, \bar D_\da] \hP \fdq{}{\hP} -
  H^{\be\db} \{D_\be, \bar D_\db\} \{ D_\al, \bar D_\da \} \hP \fdq{}{\hP}
\right) \nonumber \\[1ex]
& \quad - \tfr{1}{8} \bar D^2 \left( H_{\al\da} D^2 \bar D^\da \hP
  \fdq{}{\hP} \right) - \quar \bar D_\db D_\be \left( H^{\be\db} \bar D^2
  D_\al \hP \fdq{}{\hP} \right)
\end{align}

\noindent{\bf Weyl Transformations}
\begin{align}
w\is(A_\pm) &= -A_\pm\fdq{}{A_\pm} \\[1ex]
w\is(J) &= \fdq{}{J} \\[1ex]
w\is(H)&=0 \\[1ex]
w^\is (\hP) &= w^\is (\hP)^{(0)}+ w^\is (\hP)^{(1)} + O(H^2)\\[1ex]
 w^\is (\hP)^{(0)} &= \bar D^2 \left( \hP \fdq{}{\hP} \right) \\[1ex]
w^\is (\hP)^{(1)} &= \half \bar D^2 \{ D_\al, \bar D_\da \} \left(
 H^{\al\da} \hP \fdq{}{\hP} \right) + \half \bar D^2 \left( H^{\al\da}
 [D_\al, \bar D_\da] \hP \fdq{}{\hP} \right) \nonumber \\[1ex]
& \quad - \half \bar D^2 \left( \hP [D_\al, \bar D_\da] \left( H^{\al\da}
 \fdq{}{\hP} \right)\right)
\end{align}
{\bf Gauge Transformations}
\begin{align}
w^{(\lambda)}(A) &= i g \left( A_- \fdq{}{A_-} - A_+
  \fdq{}{A_+}\right)\\[1ex]
w^{(\lambda)}(J) &= 0 \\[1ex]
w^{(\lambda)}(H) &= 0 \\[1ex]
w^{(\lambda)}(\hP) &= i \bar D^2 \fdq{}{\hP}
\end{align}

\end{document}